\documentclass[preprint,aps,prd,amsmath,groupedaddress]{revtex4}
\usepackage{graphicx}

\def\rh{\rho}

\def\si{\sigma}

\def\ph{\phi}

\def\ps{\psi}

\def\De{\Delta}

\def\frac#1#2{\textstyle{{{#1} \over {#2}}}}

\def\vev#1{\langle {#1}\rangle}

\def\lsim{\mathrel{\rlap{\lower4pt\hbox{\hskip1pt$\sim$}}
    \raise1pt\hbox{$<$}}}
\def\gsim{\mathrel{\rlap{\lower4pt\hbox{\hskip1pt$\sim$}}
    \raise1pt\hbox{$>$}}}

\def\etal {{\it et al.}}

\newcommand{\beq}{\begin{equation}}
\newcommand{\eeq}{\end{equation}}
\newcommand{\bea}{\begin{eqnarray}}
\newcommand{\eea}{\end{eqnarray}}
\newcommand{\bse}{\begin{subequations}}
\newcommand{\ese}{\end{subequations}}
\newcommand{\rf}[1]{(\ref{#1})}

\def\nh^#1{{\hat N}^{#1}}

\def\Leff{L_{\rm eff}}
\def\Pieff{\Pi_{\rm eff}}
\def\psieff{\psi_{\rm eff}}

\def\ke{\tilde\kappa_{e+}}
\def\ko{\tilde\kappa_{o-}}

\begin{document}

\title{Sensitive polarimetric search for relativity violations 
in gamma-ray bursts}

\author{V.\ Alan Kosteleck\'y$^1$ and Matthew Mewes$^2$}
\address{$^1$Physics Department,\\
Indiana University,\\
Bloomington, IN 47405, U.S.A.\\
$^2$Physics Department,\\
Marquette University,\\
Milwaukee, WI 53201, U.S.A.\\
}

\date{IUHET 492, submitted for publication June 13, 2006\\
\vskip 0.5 truein
}

\begin{abstract}
{\bigskip
\noindent
We show that the recent measurements of linear polarization  
in gamma rays from GRB 930131 and GRB 960924 
constrain certain types of relativity violations in photons 
to less than parts in $10^{37}$,
representing an improvement in sensitivity
by a factor of 100,000. 
}
\end{abstract}

\maketitle 

At the fundamental level, 
nature is well described by
the Standard Model of particle physics
and Einstein's general relativity.
Self-consistency and aesthetics of these theories 
suggest that they coalesce into a single underlying unified theory
at extreme high energies.
Experimental evidence of this notion
is difficult to obtain because 
the energies involved are inaccessible.
However, 
minuscule effects of the coalescence might be detectable 
in suitably sensitive experiments.

A promising candidate effect is relativity violations,
which are associated with the breaking of Lorentz symmetry,
the invariance of the laws of physics under rotations and boosts
\cite{cpt04}.
Here,
we show that the recent measurement of linear polarization
in gamma rays from the gamma-ray bursts
GRB 930131 and GRB 960924 \cite{wbbcdmmss}
places a conservative bound on some types of relativity violations
in photons at the level of $10^{-37}$.
This sensitivity improves 
by a factor of 100,000 over previous analyses
\cite{km},
which used comparatively low-energy photons 
in the infrared to ultraviolet range,
and it places strong limits on relativity-violating models.

The effects of Lorentz violations at attainable energies 
are determined by a theoretical framework 
known as the Standard-Model Extension (SME),
which describes general violations of Lorentz symmetry 
independent of their 
origin \cite{ck,ak}.
The SME provides a basis for experimental searches 
for relativity violations in many different arenas,
including modern Michelson-Morley and Kennedy-Thorndike
tests highly sensitive to the properties of 
light \cite{lipa,muller,stanwix,antonini}.
In general, 
light is composed of two independently
propagating constituent waves,
each possessing a polarization and a velocity.
The SME predicts that certain forms of relativity violations
cause light to experience birefringence,
a change in polarization as it propagates 
in otherwise empty space.
The changes grow linearly with distance travelled,
so birefringence over cosmological scales
offers an exceptionally sensitive signal 
for relativity violations.
Searches for this vacuum birefringence
using near-optical polarized light
emitted from galaxies at cosmological distances
yield some of the sharpest existing tests of
relativity \cite{cfj,km}.

The SME coefficients that control Lorentz violation 
can be ordered according to their dimensionality,
which offers a simple measure of their expected size
\cite{kp}.
Among the dominant terms in the photon sector of the SME 
are ten that induce vacuum birefringence of light.
They predict changes in polarization 
that are reciprocally related to wavelength,
so using higher-energy photons increases sensitivity 
to the effects.
The corresponding ten coefficients controlling the birefringence  
are conveniently collected in matrices $\ke$ and $\ko$.
Their magnitudes are currently bounded at the level of $10^{-32}$
by observations of polarized light
from galaxies at cosmological distances
(distances $L\sim 1 {~\rm Gpc} \sim 10^{25} {~\rm m}$)
in the near optical range
(energies $E\sim 1$ eV) \cite{km}.
These sensitivities at present are
roughly comparable to the best attained values 
in matter-based relativity tests involving first-generation particles
\cite{bear,wolf,heckel,kl}
and exceed present results from interferometric studies 
with second- and third-generation particles
\cite{mesonexpt,nuexpt}.
However,
a gamma-ray burst (GRB) contains photons in the MeV range,
so GRB polarimetry offers the potential 
for a millionfold improvement in sensitivity
assuming similar propagation distances.

GRB polarimetry is presently in its infancy
\cite{wbbcdmmss,cb,rf,bc,whagz,pogo,compton,polar,estremo}.
Some time ago,
evidence was offered 
for a high degree of linear polarization $\Pi$
in emissions from GRB 021206 
extending into the MeV range \cite{cb}.
The claimed value $\Pi_{021206}=(80\pm20)\%$
was based on a study of detector scatterings of gamma-ray photons
in the Reuven Ramaty High Energy Solar Spectroscopic Imager (RHESSI).
This result has proved controversial \cite{rf,bc,whagz}
but served to stimulate interest in the subject.
Recently,
nonzero linear polarizations 
of $\Pi_{930131} >35\%$ in GRB 930131 
and of $\Pi_{960924} >50\%$ in GRB 960924 
have been demonstrated \cite{wbbcdmmss}.
These latter results are based on 
an analysis of emissions at energies up to 100 keV,
recorded by the Burst And Transient Source Experiment (BATSE)
via the scattering of gamma-ray photons off the Earth's atmosphere.

Our goal here
is to consider the implications of these observations
for the dominant relativity violations in the SME,
controlled by the coefficients $\ke$ and $\ko$.
We disregard possible gravitational couplings
\cite{ak,baik}
and assume the coefficients $\ke$ and $\ko$
are constant in spacetime.
If these coefficients originate in spontaneous Lorentz violation
\cite{ksp},
they have companion Nambu-Goldstone fluctuations
that might be interpreted as the photon \cite{bkng} 
or graviton \cite{kpng} fields,
but this is a secondary issue
and is disregarded in this work.
Subdominant birefringent effects,
which grow quadratically or faster in energy
and correspond to nonrenormalizable terms in the SME,
have been considered for GRB 021206 
in the context of specific Lorentz-violating scenarios
using modified dispersion relations \cite{qgp1,qgp2}.
Other dominant birefringent effects in the SME 
also exist \cite{cfj,ck},
but they are energy independent
and so cannot benefit from the higher photon energies 
accessible in gamma-ray bursts.

Nonzero values of the coefficients $\ke$ and $\ko$ 
modify the usual photon phase velocity 
to the form $v_\pm/c=1+\rh\pm\si$,
where $\rh$ and $\si$ are source-dependent combinations
of $\ke$ and $\ko$ \cite{ck}.
The two choices of sign,
$v_+$ and $v_-$,
correspond to the velocities 
of the two constituent birefringent waves,
which have linear and mutually orthogonal polarizations.
The difference in phase velocities induces 
a phase shift $\De\ph$ between the two component waves,
given by
\beq
\De\ph\simeq 2 \si \Leff E,
\label{Dphi}
\eeq
where $E$ is the photon energy 
and $\Leff$ is the effective light travel distance,
including corrections stemming from the expansion of the Universe
during the 
propagation \cite{km}.
We see that $\De\ph$ depends linearly 
on the combination $\si$ of coefficients for relativity violations. 
In contrast,
the combination $\rh$ affects both wave components 
equally and so induces no birefringence.

The interference between the two birefringent components
causes the polarization of light to vary 
as $\De\phi$ changes.
Through Eq.\ \rf{Dphi},
the polarization becomes dependent 
both on the effective distance $\Leff$ 
and on the photon energy $E$.
Larger values of $\Leff$ and $E$ therefore result
in increased sensitivity to the relativity-violating 
quantity $\si$.

An elegant visualization of
the effects of $\si$ on the polarization of light
is offered by the notion of the Stokes vector 
and the Poincar\'e sphere \cite{km}.
The Stokes vector $\vec s$ consists of three numbers
$\vec s =(s_1,s_2,s_3)$
forming a three-dimensional vector 
in the abstract Stokes-parameter space.
For coherent light,
$\vec s$ uniquely determines the relative degree 
of linear and circular polarization,
the polarization angle $\ps$,
and the helicity.

Figure 1 illustrates
the connection between the Stokes vector
and the standard elliptical-polarization angles
$\psi$ and $\chi$.
For the usual polarization ellipse,
$\psi$ characterizes the orientation
while $\chi$ characterizes the eccentricity and the helicity.
For the Poincar\'e sphere,
each surface point represents a unique polarization
and corresponds to the tip of a unit-normalized Stokes vector.
Pure-linear polarizations lie on the equator with $\chi=0$,
while right- and left-handed circular polarizations
are represented by the north pole
$\chi=+45^\circ$
and the south pole 
$\chi=-45^\circ$,
respectively.
Arbitrary right- and left-handed elliptical polarizations 
correspond to points on the northern and southern hemispheres.

\begin{figure}[t]
\centering
\includegraphics[width=0.4\columnwidth]{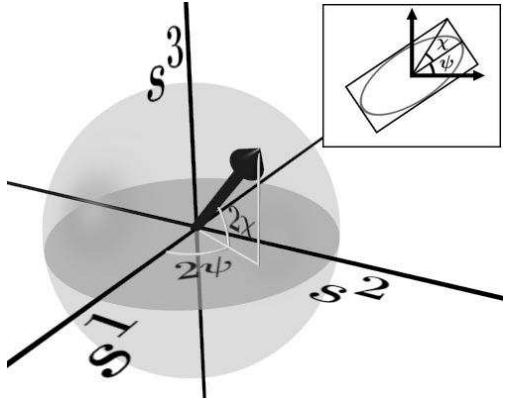}
\caption{
Elliptical polarization and the Poincar\'e sphere.
}
\label{fig1}
\end{figure}

As the light propagates
in the presence of relativity violations,
the change in polarization 
corresponds to a change in orientation of the Stokes vector.
The tip of the Stokes vector therefore follows a path 
on the Poincar\'e sphere.
Given nonzero values of $\ke$ and $\ko$, 
this path is obtained by rotating the Stokes vector
through the angle $\De\phi$ 
about an axis $\vec s_+$.
The axis $\vec s_+$ turns out to be just 
the Stokes vector for the $v_+$ component of the light alone.
Since this component is linearly polarized,
its Stokes vector $\vec s_+$ lies in the $s_1$-$s_2$ plane.
The Stokes vector $\vec s_-$ for the $v_-$ component
points in the opposite direction,
$\vec s_-=-\vec s_+$.

For a fixed photon energy $E$ 
and effective travel distance $L_{\rm eff}$,
Eq.\ \rf{Dphi} shows that the magnitude 
$\De\ph$ of the rotation,
and hence the size of the effect,
is controlled by the relativity-violating quantity $\si$.
The orientation $\vec s_+$ of the rotation axis 
also plays a role.
It can be parameterized 
by a single source-dependent angle $\xi$,
which is the angle between $\vec s_+$ and the $s^1$ axis.
For a given source,
it follows that birefringence 
and the subsequent change in polarization 
are then completely determined by $\si$ and $\xi$.

In applying these results to measurements of GRB polarization,
the key idea is that a nonzero $\si$ causes 
the polarization to change rapidly with energy 
at some large effective distance $\Leff$.
The observational signal 
is a reduced measured degree of linear polarization $\Pi$.
Two effects contribute to the reduction.
One is an energy-dependent mixing from linear to circular polarization, 
corresponding to a rotation of the Stokes vector
away from the equator of the Poincar\'e sphere.
The second is an energy-dependent change
in the polarization angle $\psi$.
This arises because as $\vec s$ rotates about $\vec s_+$
the angle $\ps$ oscillates about a mean value 
of $\xi/2$ or $(\xi+\pi)/2$.
Both these effects result in a loss of coherence over a range of $E$,
thereby reducing the measured value of $\Pi$. 

To get conservative bounds on Lorentz violation, 
we assume the emissions associated with 
GRB 930131, GRB 960924, and GRB 021206 
initially have 100\% linear polarization 
at a constant polarization angle $\psi_0$ 
over the relevant energy range 
(31-98 keV for GRB 930131 and GRB 960924 \cite{wbbcdmmss},
and 150-2000 keV for GRB 021206 \cite{cb}). 
Using simulations for the photons and an appropriate smearing
over the relevant energy range, 
we then calculate the effective degree 
of linear polarization $\Pieff$
and the effective polarization angle $\psieff$ seen at the Earth,
as a function of the relativity-violating parameters $\si$ and $\xi$.

For the simulations,
the effective distance $\Leff$
to each GRB is required.
This includes a measure of the increase in photon wavelength
resulting from the cosmological expansion.
For a universe with equation of state $p/\rho=w$ constant, 
the effective distance is then 
$\Leff=2[1-(1+z)^{-(1+3w)/2}]/(1+3w)H_0$.
At small redshifts,
this is independent of $w$:
$L_{\rm eff}=z/H_0$. 
In what follows, we use this result 
with $H_0=71$ km s$^{-1}$ Mpc$^{-1}$.

Reliable GRB redshifts are typically obtained 
through observations of their optical counterparts. 
For GRB 021206,
we take $z=0.3$ \cite{qgv,rs}.
No direct measurement is available for GRB 930131 or GRB 960924,
but there are several attempts at distance determinations 
from gamma-ray spectral data alone \cite{rs,rs2,rs3,rs4}.
For example, 
`pseudo-redshifts' of $\hat z\simeq 0.2$ are obtained 
for both GRBs applying the method of Ref.\ \cite{rs}. 
It is estimated this method is accurate to about a factor of two, 
so we take a conservative value of $z=0.1$, 
bearing in mind that the redshifts may be as high as $z=0.4$. 
The latter value would improve our sensitivity by 
about a factor of 4.

For each of GRB 930131 and GRB 960924,
we simulate albedo photons in the relevant energy range,
approximating the albedo energy spectra 
by a polynomial fit to Fig.\ 5 of Ref.\ \cite{wbbcdmmss}. 
The degree of linear polarization 
in the incident radiation is taken as 
$\Pi=\vev{s_1}^2+\vev{s_2}^2$, 
where $\vev{s_j}$ is the average of the photon Stokes parameter $s_j$. 
Drawing from the above distributions at fixed values 
of $\si$ and $\xi$, 
we generate 10,000 linearly polarized photons for each GRB 
with polarization angle $\psi_0$. 
Using Eq.\ \rf{Dphi}, 
we calculate the rotation of the Stokes vector 
during propagation 
and the resulting polarization when the photon 
reaches the Earth,
and we average over the 10,000 photons
to obtain $\Pieff$ and $\psieff$.
A more sophisticated analysis using the recorded events 
and taking into account detector efficiencies 
could in principle be performed
but is unlikely to change significantly the reported results.

For GRB 021206,
we estimate and subtract background
using all RHESSI events \cite{rhessi}
and obtain a fit to the remaining energy spectrum.
Using this fit to simulate the burst photons
and drawing from the resulting distribution 
at fixed values of $\si$ and $\xi$,
we generate a set of linearly polarized
photons with polarization angle $\psi_0$
and propagate them to the Earth via Eq.\ \rf{Dphi}. 
To model the detection process,
we simulate the scatter angle $\ph$
by drawing from the resulting distribution
and using the expression 
$dS/d\ph \propto [1-\mu\Pi\cos2(\ph-\ps)]$
for the azimuthal scatter rate,
where $\mu\simeq 0.2$ is 
the detector-dependent modulation factor.
The effective photon degree of linear polarization $\Pieff$
and the effective polarization angle $\psieff$ at the Earth
can then be extracted by fitting to 
the effective azimuthal scatter rate. 

\begin{figure}
\centering
\includegraphics[width=0.35\columnwidth]{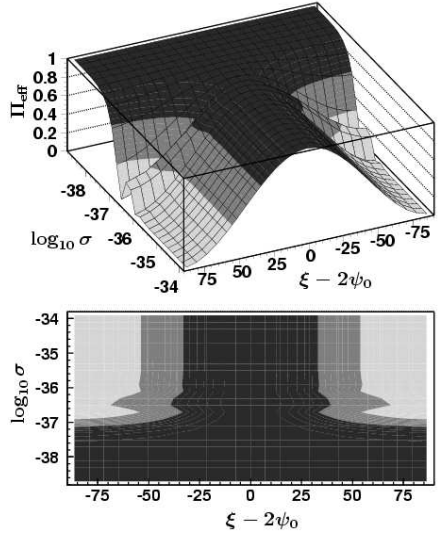}
\caption{
Effective degree of linear polarization $\Pieff$ 
as a function of $\log_{10} \si$ and $\xi$
for GRB 930131. 
$\Pieff=0\%$ - $35\%$ 
(light),
$\Pieff=35\%$ - $70\%$ 
(gray),
$\Pieff=70\%$ - $100\%$ 
(dark).
}
\label{fig2}
\end{figure}

\begin{figure}
\centering
\includegraphics[width=0.35\columnwidth]{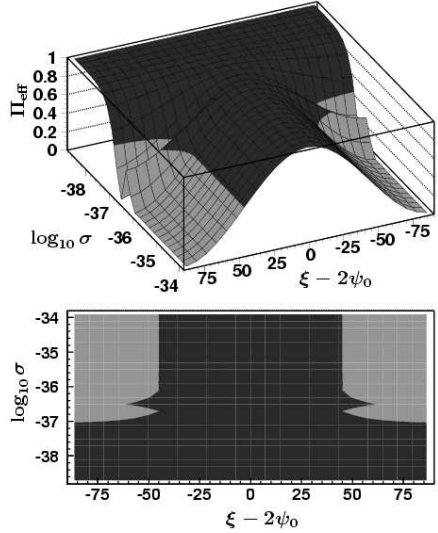}
\caption{
Effective degree of linear polarization $\Pieff$ 
as a function of $\log_{10} \si$ and $\xi$
for GRB 960924. 
$\Pieff=0\%$ - $50\%$
(gray),
$\Pieff=50\%$ - $100\%$
(dark).
}
\label{fig3}
\end{figure}

\begin{figure}
\centering
\includegraphics[width=0.35\columnwidth]{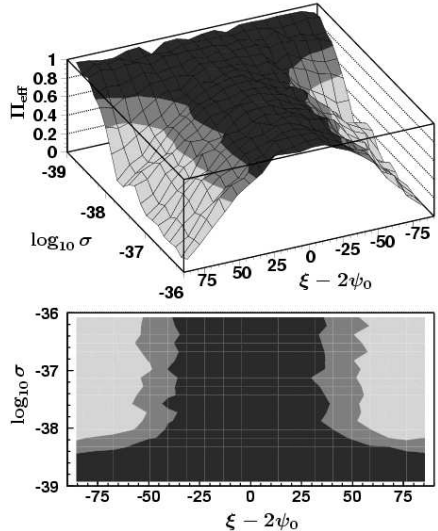}
\caption{
Effective degree of linear polarization $\Pieff$ 
as a function of $\log_{10} \si$ and $\xi$
for GRB 021206. 
$\Pieff=0\%$ - $60\%$ 
(light),
$\Pieff=60\%$ - $80\%$ 
(gray),
$\Pieff=80\%$ - $100\%$
(dark).
}
\label{fig4}
\end{figure}

Figure 2 displays the resulting $\Pieff$
as a function of $\si$ and $\xi$
for GRB 930131.
The value of $\Pieff$ lies above 70\% 
for $\si\ll 10^{-37}$ and around $\xi=2\psi_0$.
The latter behavior arises because for these values of $\xi$
the Stokes vector $\vec s$ 
is parallel to either $\vec s_+$ or $\vec s_-$,
so polarization measurements are unaffected by relativity violations.
The measurement of $\Pi_{930131} > 35\%$ \cite{wbbcdmmss}
excludes the region 
extending down to $\log_{10}\si < -37$
for this source.
Figure 3 shows the results for GRB 960924.
A similar result is obtained:
the measurement of $\Pi_{930131} > 50\%$ \cite{wbbcdmmss}
excludes the region
extending down to $\log_{10}\si < -37$.
Figure 4 displays the result for GRB 021206.
We find that the claimed value
of $\Pi_{021206}=(80\pm20)\%$ \cite{cb} 
excludes a region extending down to $\log_{10}\si < -38$.

The analysis presented here
demonstrates explicitly that the observation 
of high degrees of polarization 
for GRB photons with energies ranging to 1 MeV 
has the potential to probe the coefficients $\ke$ and $\ko$ 
at the level of $10^{-38}$.
Our results provide the most sensitive test 
for nonzero $\ke$ and $\ko$ to date,
attaining the level of $10^{-37}$.
They represent an improvement over existing sensitivities
by a conservative factor of 100,000.

This work was supported in part
by the United States Department of Energy
and by the National Aeronautics and Space Administration.

\end{document}